\documentclass[preprint,review,12pt,a4paper]{elsarticle}
\usepackage{graphics}

\usepackage{amssymb}

\biboptions{sort&compress}

\begin{document}

%
\newcommand{\IJMPB}{Int. J. Mod. Phys. B }
\newcommand{\PC}{Physica C }
\newcommand{\PB}{Physica B }
\newcommand{\JS}{J. Supercond. }
\newcommand{\IEEEmw}{IEEE Trans. Microwave Theory Tech. }
\newcommand{\IEEEas}{IEEE Trans. Appl. Supercond. }
\newcommand{\IEEEim}{IEEE Trans. Instr. Meas. }
\newcommand{\PRB}{Phys. Rev. B }
\newcommand{\PRL}{Phys. Rev. Lett. }
\newcommand{\PR}{Phys. Rev. }
\newcommand{\PL}{Phys. Lett. }
\newcommand{\IJIMW}{Int. J. Infrared Millim. Waves }
\newcommand{\APL}{Appl. Phys. Lett. }
\newcommand{\JAP}{J. Appl. Phys. }
\newcommand{\JPCM}{J. Phys.: Condens. Matter }
\newcommand{\JPCS}{J. Phys. Chem. Solids }
\newcommand{\AdP}{Adv. Phys. }
\newcommand{\Nat}{Nature }
\newcommand{\CM}{cond-mat/ }
\newcommand{\JpnJAP}{Jpn. J. Appl. Phys. }
\newcommand{\PhT}{Phys. Today }
\newcommand{\ZETF}{Zh. Eksperim. i. Teor. Fiz. }
\newcommand{\JETP}{Soviet Phys.--JETP }
\newcommand{\EL}{Europhys. Lett. }
\newcommand{\Sci}{Science }
\newcommand{\EJPB}{Eur. J. Phys. B }
\newcommand{\IJMB}{Int. J. of Mod. Phys. B }
\newcommand{\RPP}{Rep. Prog. Phys. }
\newcommand{\SUST}{Supercond. Sci. Technol. }
\newcommand{\JLTP}{J. Low Temp. Phys. }
\newcommand{\RSI}{Rev. Sci. Instrum. }
\newcommand{\RMP}{Rev. Mod. Phys. }
\newcommand{\LTP}{Low Temp. Phys. }
\newcommand{\MST}{Meas. Sci. Technol. }

\begin{frontmatter}

\title{\normalsize
BL-6-INV / ISS2010$\;\;\;\;\;\;\;\;\;\;\;\;\;\;\;\;\;\;\;\;\;\;\;\;\;\;\;\;\;\;\;\;\;\;\;\;\;\;\;\;\;\;\;\;\;\;\;\;\;\;\;\;\;\;\;\;\;\;\;\;\;\;\;\;\;\;\;\;\;\;\;\;\;\;\;\;\;\;\;\;\;\;\;\;\;\;\;\;\;\;\;\;\;\;\;\;$\\
Submitted November 2, 
2010$\;\;\;\;\;\;\;\;\;\;\;\;\;\;\;\;\;\;\;\;\;\;\;\;\;\;\;\;\;\;\;\;\;\;\;\;\;\;\;\;\;\;\;\;\;\;\;\;\;\;\;\;\;\;\;\;\;\;\;\;\;\;\;\;\;\;\;\;\;\;\;\;\;\;\;\;\;\;\;\;\;\;$\\
\vspace{1cm}
\Large
Microwave properties of DyBCO monodomain in the mixed state and 
comparison with other RE-BCO systems}

\author[Roma3]{N. Pompeo$^*$}
\author[Roma3]{R. Rogai}

\author[B1]{M. Ausloos}
\author[B2]{R. Cloots}

\author[Enea]{A. Augieri}
\author[Enea]{G. Celentano}

\author[Roma3]{E. Silva}

\address[Roma3]{Dipartimento di Fisica "E. Amaldi" and CNISM, 
Universit\`a Roma Tre,
Via della Vasca Navale 84, 00146 Rome, Italy}

\address[B1]{SUPRATECS, Sart-Tilman, B-4000 Liege, Belgium}
\address[B2]{LSIC, Chemistry Department B6, University of Liege, 
Sart-Tilman, B-4000 Liege, Belgium}

\address[Enea]{Associazione EURATOM-ENEA, UT Fusione - 
Superconductivity Laboratory, Centro Ricerche Frascati,
Via E. Fermi 45, 00044 Frascati (Rome), Italy}

\begin{abstract}
We report on microwave measurements on DyBa$_2$Cu$_3$O$_{7-\rm\delta}$ 
monodomains grown by the  top-seeded melt-textured technique. We 
measured the field increase of the surface resistance $R_{\rm s}(H)$ in the 
a-b plane at 48.3 GHz. Measurements were performed at fixed 
temperatures in the range 70 K - $T_{\rm c}$ with a static magnetic field 
$\mu_0H<0.8$ T parallel to the c-axis.
Low field steep increase of the dissipation, typical signature of the 
presence of weak links, is absent, thus indicating the single-domain 
behaviour of the sample under study. The magnetic field dependence of 
$R_{\rm s}(H)$ is ascribed to the dissipation caused by vortex motion. The 
analysis of $X_{\rm s}(H)$ points to a free-flow regime, thus allowing to 
obtain the vortex viscosity as a function of temperature. We compare 
the results with those obtained on RE-BCO systems. In particular, we 
consider strongly pinned films of YBa$_2$Cu$_3$O$_{7-\rm\delta}$ with 
nanometric BaZrO$_3$ inclusions.\\
\vspace{1cm}

\vspace{0.1cm}

{
\noindent Corresponding author:\\
Dr. Nicola Pompeo\\
Postal address: Dipartimento di Fisica, Universit\`a Roma Tre, Via della Vasca Navale 84 - 00146 Rome, Italy\\
Phone: +39 06 57337260\\
Fax: +39 06 57337102\\
E-mail address: pompeo@fis.uniroma3.it
}

\end{abstract}




\end{frontmatter}

\newpage
\section{Introduction}
The microwave electrodynamic response in High-$T_{\rm c}$ Superconductors 
(HTCS) is a precious tool in the investigation of these materials. It 
provides a great deal of information concerning fundamental physics 
\cite{narlikar}, as well as allowing to address essential issues in 
view of technological applications \cite{gallop}.
The microwave response determined at zero field has allowed to 
address many points such as the temperature dependence of the 
superfluid density \cite{hardyPRL93,hosseiniPRL98,maedaJPCM05}, and 
the quasi-particles (QP) properties above and below the 
superconducting transition 
\cite{maedaJPCM05,hosseiniPRL98,hosseiniPRB99}.
By applying a static magnetic field $H>H_{\rm c1}$, HTCS are driven in 
the mixed state where the presence of vortices allows for the 
disclosure of additional physics \cite{maedaJPCM05}. Vortices, which 
are set in motion by the Lorentz force exerted by microwave currents, 
dissipate energy through the QP excitations located in and around 
their cores, in which the order parameter is depressed. Because of 
the nature of their cores, vortices can be considered as a window of 
``quasi-normal'' state properties accessible below $T_{\rm c}$, embedded in 
the superconducting condensate, and thus useful to probe 
``normal''-state-related properties, simultaneously with the 
superconducting gap issues (in particular, its symmetry).

 From the point of view of  fundamental physics, it is then 
interesting to investigate the vortex dissipation, dictated by the 
quasi-particle density of states (DOS) and relaxation time in the 
vortex cores.

At the same time, from a technological point of view, it is well 
known \cite{blatterone} that the power handling of HTCS, relevant to 
microwave devices, is limited by grain-boundaries contribution 
(dominant at low fields) as well as by vortex motion, the latter 
being the ultimate, unavoidable limitation. Within this scenario, the 
investigation of vortex pinning mechanisms is an essential task.
Single crystals are ideal systems for the study of intrinsic 
properties, while epitaxial films are of interest for applications. 
On the other hand, monodomains, despite their technological interest, 
are rarely the subject of microwave studies.
Therefore, in this paper we will present the microwave 
characterization and study of DyBa$_{2}$Cu$_{3}$O$_{7-\rm\delta}$ 
(DyBCO) monodomains. A very few studies of DyBCO at microwaves in the 
mixed state exist \cite{altriDyBCO}, while the parent compound 
YBa$_{2}$Cu$_{3}$O$_{7-\rm\delta}$ (YBCO) is widely studied. It will 
demonstrate particularly useful a comparison between data taken in 
DyBCO monodomains, in YBCO single crystals \cite{tsuchiyaPRB01} and 
in YBCO epitaxial thin films with artificially enhanced pinning 
\cite{pompeoAPL91}, as prototypical case for intrinsic and extrinsic 
behaviour, respectively.\\

\section{Experimental technique and data analysis}
\label{sec:technique}
The main experimental quantity in microwave experiments is the 
effective surface impedance $Z_{\rm s}=R_{\rm s}+{\rm i}X_{\rm s}$. In this work, the 
surface impedance is measured by means of two cylindrical resonators, 
a silver-coated metal cavity \cite{silvaCavity} and a dielectric 
resonator \cite{dielRes}, operating in the TE$_{011}$ mode at 
approximately 48.3 GHz and 47.7 GHz, respectively. The surface 
perturbation method is used, with the sample under measurement 
replacing one of the cavity bases. The microwave currents flow 
parallel to the sample surface (along the $a-b$ planes for the 
$c$-axis oriented samples) on circular patterns. A solid/liquid 
nitrogen cryostat allows to reach temperatures $T$ as low as 60 K, 
with temperature control within $\pm$0.005 K. A conventional 
electromagnet generates magnetic fields $\mu_0H\leq0.8$ T, applied 
perpendicularly to the probed surface of the sample (i.e. parallel to 
the superconductor $c$-axis in $c$-axis oriented samples).
The field-dependent cavity quality factor $Q$ and resonant 
frequency $\nu$ are measured to yield the corresponding surface 
impedance  variations according to the following equations:
\begin{eqnarray}
\label{eq:ZfromQ}
\Delta R_{\rm s} (H,T)&=&R_{\rm s} (H,T)-R_{\rm s} (0,T)=\nonumber\\
&=&G\left[\frac{1}{Q(H,T)}-\frac{1}{Q(0,T)}\right]\\
\Delta X_{\rm s} (H,T)&=&X_{\rm s} (H,T)-X_{\rm s} (0,T)=\nonumber\\
&=&-2G\frac{\nu(H,T)-\nu(0,T)}{\nu(0,T)}
\end{eqnarray}

\noindent where $G$ is a geometric factor of the cavity which can be 
computed from the theoretically known distribution of the 
electromagnetic field. Here, $G\approx10840$ and $G\approx2000$, for 
the cavity and the dielectric resonator, respectively.
Samples smaller than the base of the resonators are accommodated with 
the aid of an auxiliary thin metallic mask. In this case the 
geometric factor increases and sensitivity decreases.

Measurements are performed by quasi-statically sweeping the applied 
field intensity $H$ at fixed temperature after zero field cooling.
It is worth noting that by considering field-induced variations of 
$Z_{\rm s}$, no calibration of the cavity response is needed since the 
latter is field independent.\\
As already anticipated in the previous Section, the in-field surface 
impedance is determined by two main contributions: grain boundaries 
and vortex motion.\\
Grain boundaries, depending on the misalignment angle between 
adjacent grains, exhibit behaviours ranging from metallic to 
Josephson tunneling. In magnetic fields, they constitute preferential 
paths for the motion of vortices, yielding generally lower pinning 
forces along their direction: the actual nature of the vortices 
located in the GB depends again on the misalignment angle. With 
larger and larger misalignment angle, the nature of vortices changes 
from standard Abrikosov vortices to the so-called Abrikosov-Josephson 
(AJ) vortices, and finally to core-less Josephson vortices 
\cite{gurevichPRB65}.
Many models have been developed in order to capture the GB behaviour 
in the microwave regimes 
\cite{marconPRB39,hyltonPRB39,halbritter90,wosik93}.
Independently from the adopted model, the main signature of GB in the 
in-field microwave surface impedance consists in an abrupt, 
quasi-step-like increase of the surface resistance $R_{\rm s}$ with 
increasing field, followed by a flat plateau \cite{marconPRB39, 
heinJAP75}. The field scale over which the step actually extends 
varies from a few mT \cite{marconPRB39, heinJAP75} for weak-links and 
Josephson vortices up to 0.1 T or larger for AJ vortices 
\cite{gurevichPRB65,gurevichPRL88}.\\
Abrikosov vortex motion within intragrain regions is the ubiquitous 
phenomenon visible in surface impedance measurements in the mixed 
state.
The mixed state microwave response, which includes vortex dynamics, 
is quite intricate, since it emerges from the interplay between the 
currents excited by the applied microwave fields and vortices set in 
motion by these currents.
Many authors considered this issue, providing models which take into 
accounts various aspects \cite{GR,CC,brandt,MStheory,dulcicPC93}. 
Following Coffey-Clem (CC) approach \cite{CC}, the whole complex 
resistivity $\tilde\rho$ can be written down as follows:
\begin{equation}
\label{eq:rhoc}
     \tilde{\rho}=\frac{\rho_{\rm vm}+{\rm 
i}{\omega\mu_0\lambda^2}}{1+{\rm i}\frac{2\lambda^2}{\delta_{\rm nf}^2}}
\end{equation}
\noindent where $\omega=2\pi\nu$ is the microwave angular frequency, 
$\rho_{\rm vm}$ is the complex resistivity due to Abrikosov vortex 
motion, and $\lambda$ and $\delta_{\rm nf}$ are the London and normal 
fluid penetration depths, respectively.\\
Vortex dynamics involves many mechanisms: the interaction with the 
superconducting condensate yields a viscous drag, described through a 
viscous drag coefficient (also commonly called vortex viscosity) 
$\eta$. The interaction between crystal defects and the fluxon system 
generates a pinning effect usually described through the pinning 
constant $k_{\rm p}$, applicable in the limit of small vortex displacements 
from their equilibrium positions as determined by high frequency 
stimuli. Thermal fluctuations allow for thermally activated flux 
jumps between pinning sites, yielding the so-called creep.\\
One finds \cite{pompeoPRB78} that a large variety of different models 
can be formulated under the following  very general expression for 
the vortex resistivity $\rho_{\rm vm}$:
\begin{equation}
\label{eq:rhovm}
     \rho_{\rm vm}=\rho_{\rm ff}\frac{\epsilon+{\rm 
i}\frac{\omega}{\omega_{0}}}{1+{\rm i}\frac{\omega}{\omega_{0}}}
\end{equation}
\noindent where $\rho_{\rm ff}=\Phi_{0}B/\eta$ is the flux flow 
resistivity, $B$ the magnetic induction field, $\Phi_0$ the flux 
quantum, $\epsilon$ a dimensionless creep factor, constrained in the 
range [0,1]. When creep can be neglected ($\epsilon=0$), the above 
expression reverts to the well-known Gittleman-Rosemblum model 
\cite{GR}.\\
The relation (in the local limit) between the superconductor complex 
resistivity $\tilde\rho$ and the measured surface impedance depends 
on the penetration depth of the e.m. field with respect to the 
superconducting sample thickness $d$: for bulk samples, i.e. 
$d<<\min(\lambda, \delta_{\rm n})$, one has:
\begin{equation}
\label{eq:zbulk}
   Z_{\rm s}(H,T)=\sqrt{ {\rm i}\omega\mu_{0}\tilde{\rho}}
\end{equation}
\noindent whereas in thin films, for which the $d>>\min(\lambda, 
\delta_{\rm n})$ condition holds,\cite{thinfilm}:
\begin{equation}
\label{eq:zfilm}
   Z_{\rm s}(H,T)=\frac{\tilde\rho}{d}
\end{equation}

\section{Measurements and Discussion}
\subsection{DyBCO single domains}

DyBCO single domains were prepared with precursor powders 
DyBa$_{2}$Cu$_{3}$O$_{7-\rm \delta}$ and Dy$_{2}$BaCuO$_{5}$, produced by 
solid-state synthesis from Dy$_{2}$O$_{3}$, BaCO$_{3}$ and CuO 
powders. The powder mixture was pressed uni-axially to give 
cylindrical pellets of 10.8 mm diameter, which were melt-textured in 
atmospheric air conditions using a Nd-123 single-crystal seed. Large 
``quasi-single-crystals'', mainly $c$-axis oriented, have been 
obtained, with $T_c\sim$ 88-89 K. Two distinct pellets, (A) and (B), 
having similar $T_c\approx$ 88 K, will be considered in the following.

 From the first pellet (A), two samples of about the same thickness 
$\sim 1$ mm were cut: one (A2), approximately 2$\times$2 mm$^2$ 
square, has been characterized by a magneto-optic (MO) study.

Magneto-optic images are reported in Fig. \ref{fig:MO} (bright 
regions denote higher field intensity, black regions denote zero 
field): a static magnetic field is applied, perpendicularly to the 
sample surface, after a ZFC of the sample down to 73.5 K. In panel 
(a) the field is set to 15 $\mu$T: the sample is not threaded by 
magnetic flux, thus exhibiting a single domain behaviour. At higher 
field values (90 $\mu$T, panel (b)), the magnetic flux penetrates in 
the sample: a few cracks, along which flux lines preferentially enter 
the sample are visible. Going back to zero field (panel (c)), remnant 
flux (indicative of significant pinning) is visible with slight 
dishomogeneities, apart from the cracks, along which flux lines 
easily exit from the sample volume. Through the MO analysis the 
sample shows an overall single domain behaviour, with no visible 
microscale flux penetration.\\
The parent sample (A1) is a $\sim$10.2 mm side square and is used for 
the microwave measurements \cite{DyBCO} performed by means of the 
cavity operating at 48.3 GHz.
Measurements of $\Delta R_{\rm s}(H)$ and $\Delta X_{\rm s}(H)$ at selected 
temperatures are reported in Fig. \ref{fig:ZsDY01}.
It can be seen that the surface resistance increases with the field, 
with steeper increase at larger $T$. At low fields there is no 
evidence for step-like increase of the dissipation: we deduce that no 
significant contribution to the losses comes from weak-links or 
Josephson fluxons. The absence of signatures of JJ or AJ vortices 
confirms the MO: the sample does not present significant grain 
boundaries, thus behaving as a single domain also at microwaves.

On the other hand, the surface reactance is featureless and 
essentially flat, remaining within the same error range ($\pm$ 0.01 
$\Omega$, denoted by the thick horizontal lines) for the temperature 
range here presented. Despite large scattering than in $R_{\rm s}$ data, 
due to the sensitivity limits of the cavity used for the 
measurements, $\Delta X_{\rm s}(H)$ clearly remains well below $\Delta R_{\rm s}$ 
and constant with $H$: this is a clear indication of irrelevance of 
pinning at our measuring frequency. In fact, one can see that the 
primary effect of pinning is an increase of the reactance (connected 
to the elastic forces recalling vortices to the pinning centers). 
Analytically, it can be seen by inserting Eq. (\ref{eq:rhovm}) in Eq. 
(\ref{eq:zbulk}), and taking limits for small fields and large 
pinning, that $\Delta X_{\rm s}/\Delta R_{\rm s}\sim k_{\rm p}/(\eta\omega)$. Since we 
find that $\Delta X_{\rm s}(H)$ is negligible with respect to $\Delta 
R_{\rm s}(H)$, we can safely neglect pinning in the analysis of our data.

It is worth stressing that undetectable pinning in the high 
frequency regime does not necessarily imply small pinning in the d.c. 
transport regime: high frequencies force flux lines to undergo to 
very small oscillations (small displacement regime), thus probing 
mainly the steepness of the pinning wells. By contrast, low 
frequencies and dc force flux lines to large displacements, thus 
probing mainly the pinning barrier heights. Indeed, similar DyBCO 
samples were found to present significant pinning through d.c. and 
magnetic characterizations \cite{DYBCOmag}. The same presence of 
remnant magnetization, as seen with MO, points to a significant 
pinning in the quasistatic regime.

By neglecting pinning effects (therefore neglecting also creep), and 
by considering that, not too close to $T_{\rm c}$, the London penetration 
depth is much shorter than the normal fluid penetration depth, Eq. 
\ref{eq:rhoc} reverts to $\tilde{\rho}\approx\rho_{\rm vm}+{\rm 
i}{\omega\mu_0\lambda^2}$ with $\rho_{\rm vm}=\Phi_0 B/\eta$, which 
allows to compute the surface resistance in the bulk limit as:
\begin{equation}
\label{eq:Rsbulk}
\Delta R_{\rm s} (B,T)= \frac{\omega\mu_0\lambda}{\sqrt{2}}
\sqrt{-1+\sqrt{
1+\left(\frac{1}{\omega\mu_0\lambda^2}\frac{\Phi_{0} B}{\eta}\right)^2
}}
\end{equation}
The theoretical expression Eq. (\ref{eq:Rsbulk}) predicts a crossover 
from the linear behaviour at low fields,
\begin{equation}
\label{eq:Rsbulklin}
\Delta R_{\rm s} =\frac{1}{\lambda}\frac{\Phi_{0} B}{\eta},
\label{Rslow}
\end{equation}
when the material has a (real) flux flow resistivity 
$\rho_{\rm ff}=\frac{\Phi_{0} B}{\eta}$ and screening is dictated by 
superfluid over the London penetration depth $\lambda$, to a 
square-root dependence at higher fields. The crossover is ruled by a 
threshold field value proportional to $\lambda^2\eta$. In the general 
case, Eq. (\ref{eq:Rsbulk}) yields a downward curvature on $\Delta 
R_{\rm s}(H)$, which allows to fit well the experimental data. Within the 
London limit $B\simeq \mu_0 H$, and fits of the experimental curves 
$\Delta R_{\rm s}(H)$ can be performed. Full fits, together with the 
fitting parameters $\lambda$ and $\eta$, have been reported elsewhere 
\cite{DyBCO}. Sample fits are reported in Fig. \ref{fig:ZsDY01}. In 
the following Section \ref{sec:comparison}, we will discuss the 
viscosity values derived from fits of the data with Eq. 
\ref{eq:Rsbulk}.\\
The second sample (B1) cut from pellet (B) presents additional features.
Surface resistance data for selected temperatures are shown in Fig. 
\ref{fig:RsDY02}.
It can be seen that $\Delta R_{\rm s}(H)$ increases at a quicker rate than 
in sample A1: given their similar $T_{\rm c}$, this implies that sample B1 
is more dissipative\footnote{This fact also contributes to limiting 
the system sensitivity so that surface reactance can not be 
determined.}.
Moreover, an initial step is visible at moderately low fields (up to 
$\approx 0.1$ T), after which $\Delta R_{\rm s}$ increases linearly with 
the field, in contrast with the downward curvature seen on sample A1. 
The initial step is, as already anticipated, typical of grain 
boundaries, and the linear increase of $\Delta R_{\rm s}$ is due to 
standard Abrikosov vortex dynamics. A prototypical measure of this 
kind of behaviour is shown by data taken on granular samples, like 
the data for $\Delta R_{\rm s}$ at 23 GHz of a 
GdBa$_2$Cu$_3$O$_{7-\rm\delta}$ granular sample \cite{silvaPC218} 
reported in the inset of Fig. \ref{fig:RsDY02}.
Fits of the linear part of the data in Fig. \ref{fig:RsDY02} yield 
the product $\lambda\eta$ vs. $T$ as fit parameter. We find that 
$\lambda\eta$ is approximately the same in the two samples, as 
reported in Fig. \ref{fig:RsDY02}, panel (b). However, in sample B1 
$\Delta R_{\rm s}\propto B$, whereas in sample A1 $\Delta R_{\rm s}$ has a 
downward curvature. This fact implies that (a) $\lambda$ in sample B1 
is larger than in A1, indicating a larger penetration of the 
microwave field and that (b) $\eta$ is smaller.

An estimate of $\lambda$ and $\eta$ can be given as follows. The 
product $\lambda\eta$ is fixed by the fits of data with Eq. 
(\ref{eq:Rsbulklin}): $\lambda$ is then increased until the 
calculated curve by Eq. (\ref{eq:Rsbulk}) coincides with the data. 
This procedure yields $\lambda_{\rm(sample B1)}\geq 3.5 
\lambda_{\rm(sample A1)}$, and $\eta_{\rm(sample B1)}\leq 3.5 
\eta_{\rm(sample A1)}$. A larger $\lambda$ implies a larger 
disomogeneity in the sample, coherent with the stronger signature of 
GB observed; the simultaneous reduction of $\eta$ will be discussed 
below.

\subsection{YBCO thin films with artificially enhanced pinning}

In the previous Subsection we have presented microwave measurements 
on DyBCO single domain samples, detecting no pinning (at least, no 
significant pinning visible in this high frequency dynamic regime), 
which allowed us to extract the vortex viscosity, a physical quantity 
connected with intrinsic properties of the material.
In this Subsection we briefly present data taken on YBCO epitaxial 
thin films, where nanosize BaZrO$_3$ (BZO) particles were 
intentionally added in order to artificially improve the pinning 
properties. These systems qualify themselves as strongly 
``extrinsic'' with respect to the DyBCO monodomain. The YBCO/BZO thin 
(thickness $d\approx$ 120 nm, $T_{\rm c}$= 90 K) film here considered was 
grown by pulsed laser ablation from a target with 7 mol.\% BZO 
content. BZO particles are typically extended (correlated) defects, 
of transverse size of a few nm \cite{galluzziIEEE07}.
Measurements were performed with the dielectric resonator at 47.7 
GHz. A typical measurement at 63 K is reported in Fig. 
\ref{fig:YBCOBZO}. It is worth to stress that a large surface 
reactance is observed, pointing to a very strong pinning in this 
sample: for the measurement shown, $\Delta X_{\rm s}(H)>\Delta R_{\rm s}(H)$. 
Data analysis is straightforward: no initial steps are detected, 
indicating good connectivity of the film. Given the thickness of the 
film, Eq. \ref{eq:zfilm} holds; since $T\ll T_c$, 
$\lambda\ll\delta_{\rm n}$ and creep can be safely considered negligible, 
so that Eqs. (\ref{eq:rhoc}) and (\ref{eq:rhovm}) yield 
$\tilde\rho=\rho_{\rm ff}\frac{1}{1-{\rm i}\nu_{\rm p}/\nu}+{\rm 
i}\omega\mu_0\lambda^2$. Finally, by considering only the 
field-induced variations and neglecting any pair-breaking effects on 
$\lambda$ since $B\ll B_{c2}$, one has in the London limit $\Delta 
Z_{\rm s}(B\approx \mu_0 H)=\frac{\rho_{\rm ff}}{d}\frac{1}{1-{\rm 
i}\nu_{\rm p}/\nu}$. Therefore the quantities $\eta$ and 
$k_{\rm p}=2\pi\nu_{\rm p}/\eta$ can be directly derived from $\Delta R_{\rm s}(H)$ and 
$\Delta X_{\rm s}(H)$, without the need of any fitting procedure. A 
detailed study of the possible effect of creep has been reported in 
\cite{pompeoPRB78}. Extended reports on YBCO/BZO have been published 
elsewhere \cite{pompeoAPL91, pompeoJAP105}. We are here concerned 
specifically with the values of the vortex viscosity, in order to 
compare such intrinsic quantities in different RE-BCO systems.

\section{Comparison of vortex viscosities}
\label{sec:comparison}

We now comment and draw some conclusions about the vortex viscosities 
as extracted through mixed state microwave measurements performed on 
various materials and samples.
In Fig. \ref{fig:alleta}, viscosities of various 123 systems are 
shown: namely, the DyBCO monodomain measured at 48.3 GHz, the 
YBCO/BZO thin film measured at 47.7 GHz, and a YBCO single crystal 
measured at 40.8 GHz (from Ref. \cite{tsuchiyaPRB01}).

 From Fig. \ref{fig:alleta}, it can be seen that all the considered 
samples exhibit viscosities with similar absolute values and 
temperature dependencies. Since $\eta$ is determined by the 
quasi-particles scattering times and  density of states in the vortex 
cores (together with a contribution arising from the regions around 
the vortices, given the d-wave nature of HTCS \cite{kopnin}), this 
fact allows to infer that all these 123 systems, despite their 
structural and ``extrinsic'' differences, share the same fundamental 
physics of quasiparticles in the mixed state. This is contrasted by 
the measurements taken in DyBCO sample B1, where we estimate a vortex 
viscosity smaller by a factor $\sim 3.5$ (continuous line in Fig. 
\ref{fig:alleta}).

The latter observation deserves a specific comment. In fact, an 
explanation can be given as follows: since $\eta\propto n\tau$ (being 
$n$ and $\tau$ the quasi-particles concentration and scattering time, 
respectively) \cite{blatterone}, it comes out that sample B1 is 
affected by a significant disorder on the nanoscale (appreciable on 
scales of the order of the vortex size). The simultaneous appearance 
of disorder on microscale, as indicated by the appearance of 
weak-links fingerprints (see Fig. \ref{fig:RsDY02} and the discussion 
of the weak-links induced microwave losses in Ref. 
\cite{marconPRB39}), is an intriguing coincidence that will be 
further studied in future experiments.

\section{Conclusions}
We have presented and compared mixed state microwave measurements 
performed on various HTS RE-BCO system, including DyBCO monodomains 
and YBCO epitaxial thin films with artificially introduced 
nanoparticles. We have shown that the microwave techniques gicves 
important information, complementary to dc investigation. In 
particular, it is possible to obtain the intrinsic vortex viscosity 
even in very strongly pinned materials. We have extensively 
characterized DyBCO monodomains, and we have found that they can 
exhibit the same microwave viscosity than YBCO crystals and thin 
films with artificially included nanoparticles, thus indicating the 
same quasiparticle physics. We have also shown that DyBCO monodomains 
in some case can exhibit simultaneous microscale inhomogeneity, 
leading to weak-links, and possible nanoscale inhomogeneity, leading 
to a small scattering time as revealed by smaller viscosity.

\newpage
\section*{Acknowledgements}
We thank Samuel Denis for useful help in the preliminary stage of this 
work. This work has been partially supported by EURATOM and by an 
Italian FIRB 2008 project (SURE:ARTYST).



\newpage
\begin{figure}[htb]
\begin{center}
\caption{MO images of DyBCO sample A1, taken at 73.5 K after 
zero-field cooling, by applying a magnetic field equal to 15 
$\mu$T (upper panel), 90 $\mu$T (middle panel) and going to zero again (lower panel). Bright regions 
denote higher field intensity, black regions denote zero field.}
\label{fig:MO}
\end{center}
\end{figure}
\begin{figure}[h]
\begin{center}
\caption{Field-induced variations of the surface impedance of DyBCO 
sample A1 at 70 K and 84 K , {upper panel} and {lower 
panel}, respectively. Continuous lines are fit of $\Delta R_{\rm s}$ according to Eq. \ref{eq:Rsbulk}.}
\label{fig:ZsDY01}
\end{center}
\end{figure}
\begin{figure}[h]
\begin{center}
\caption{Upper panel: $\lambda\eta$ products for DyBCO samples A1 e B1 (bicolor and crossed squares, respectively). Lower panel: field-induced variations of the surface resistance of DyBCO 
sample B1 at selected temperatures. The straight lines are guides for the eye. Inset of lower panel: data at 23 GHz of a GdBa$_2$Cu$_3$O$_{7-\rm\delta}$ granular sample \cite{silvaPC218}.}
\label{fig:RsDY02}
\end{center}
\end{figure}
\begin{figure}[h]
\begin{center}
\caption{Field-induced variations of the surface impedance of 
YBCO/BZO thin film at 63 K.}
\label{fig:YBCOBZO}
\end{center}
\end{figure}
\begin{figure}[b]
\begin{center}
\caption{Vortex viscosity vs $T/T_c$ for a set of 123 materials 
measured at similar frequencies: DyBCO single domain (sample A1), full dots; 
YBCO/BZO thin film \cite{pompeoAPL91}, open circles; YBCO single 
crystal \cite{tsuchiyaPRB01}, squares. The continuous line is the estimated value for DyBCO sample B1.}
\label{fig:alleta}
\end{center}
\end{figure}
{ }
\clearpage
{ }
\begin{figure}[htb]
\begin{center}
\includegraphics[width=5cm]{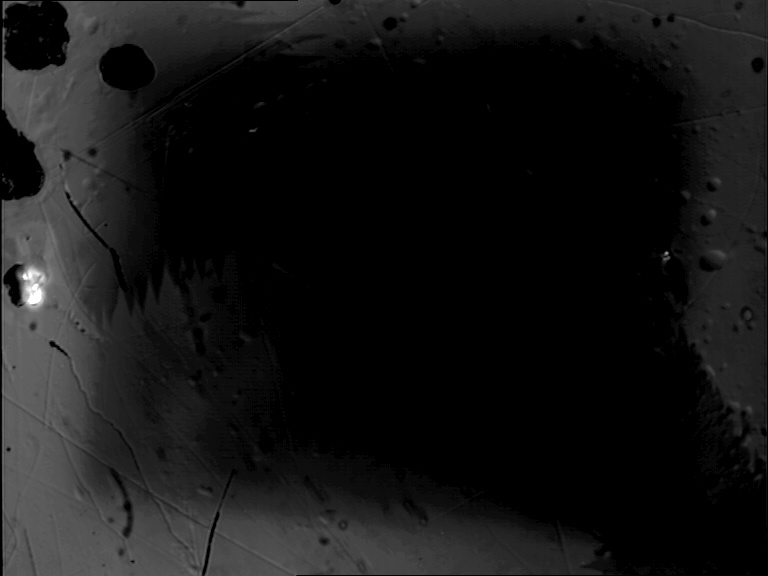}
\includegraphics[width=5cm]{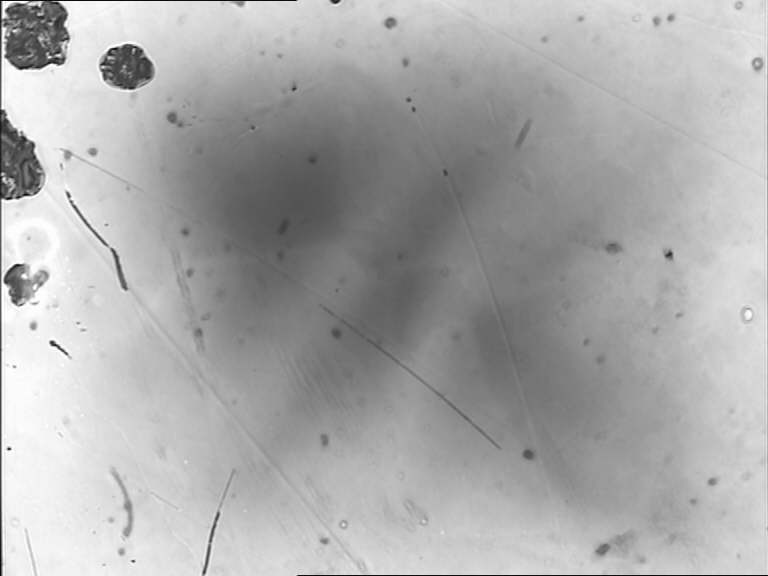}
\includegraphics[width=5cm]{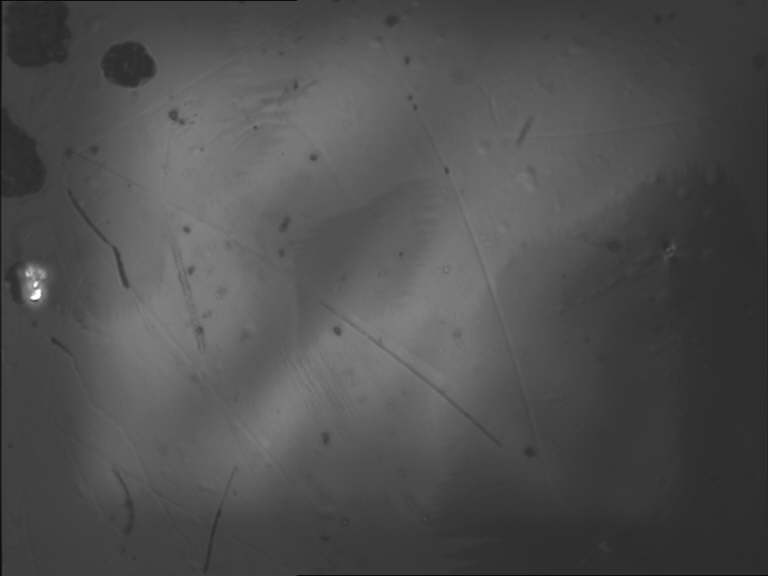}
\\
{Figure 1}
\end{center}
\end{figure}
\begin{figure}[htb]
\begin{center}
\includegraphics[width=7cm]{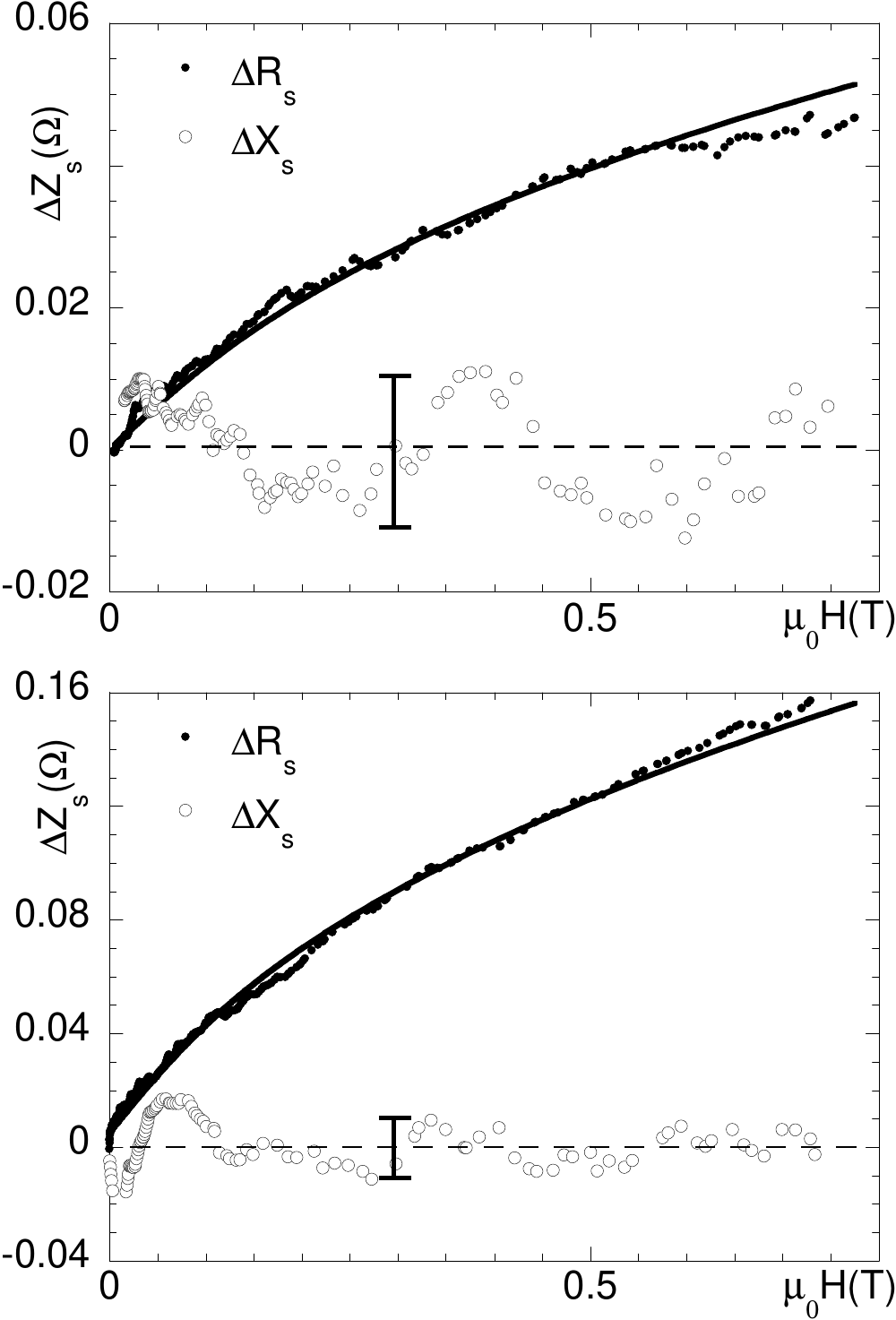}
\\
{Figure 2}
\end{center}
\end{figure}
\begin{figure}[htb]
\begin{center}
\includegraphics[width=7cm]{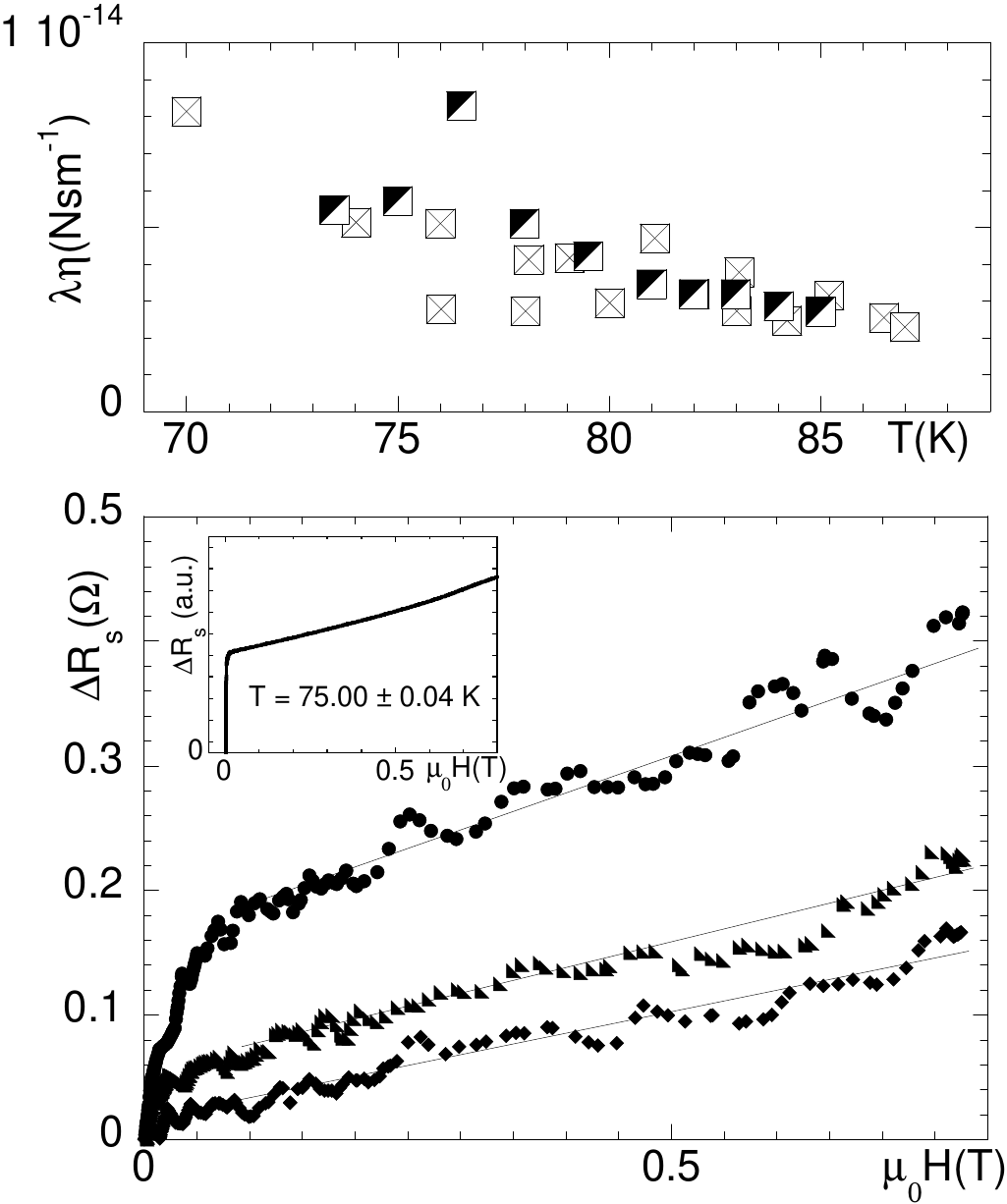}
\\
{Figure 3}
\end{center}
\end{figure}
\begin{figure}[htb]
\begin{center}
\includegraphics[width=7cm]{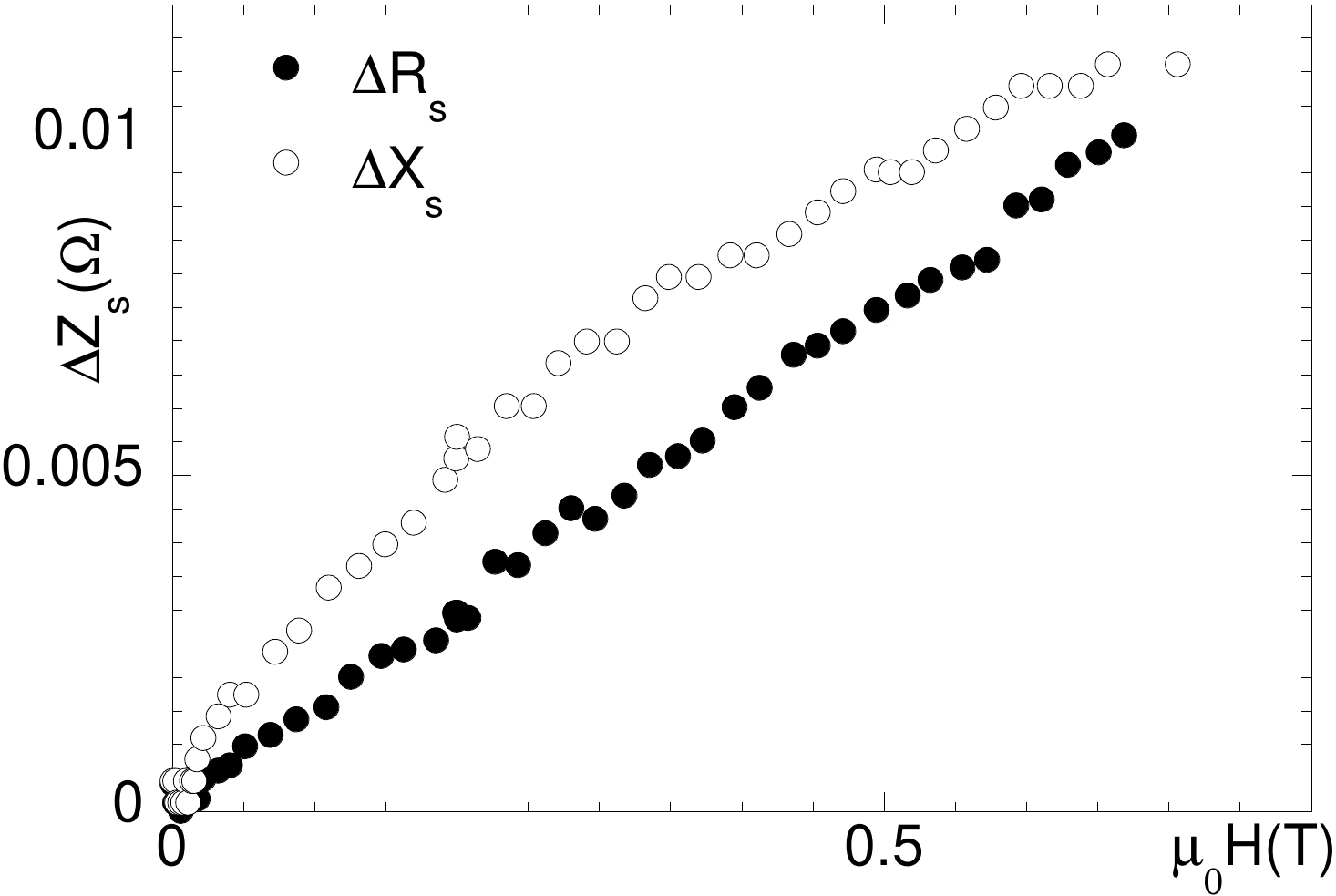}
\\
{Figure 4}
\end{center}
\end{figure}
\begin{figure}[htb]
\begin{center}
\includegraphics[width=7cm]{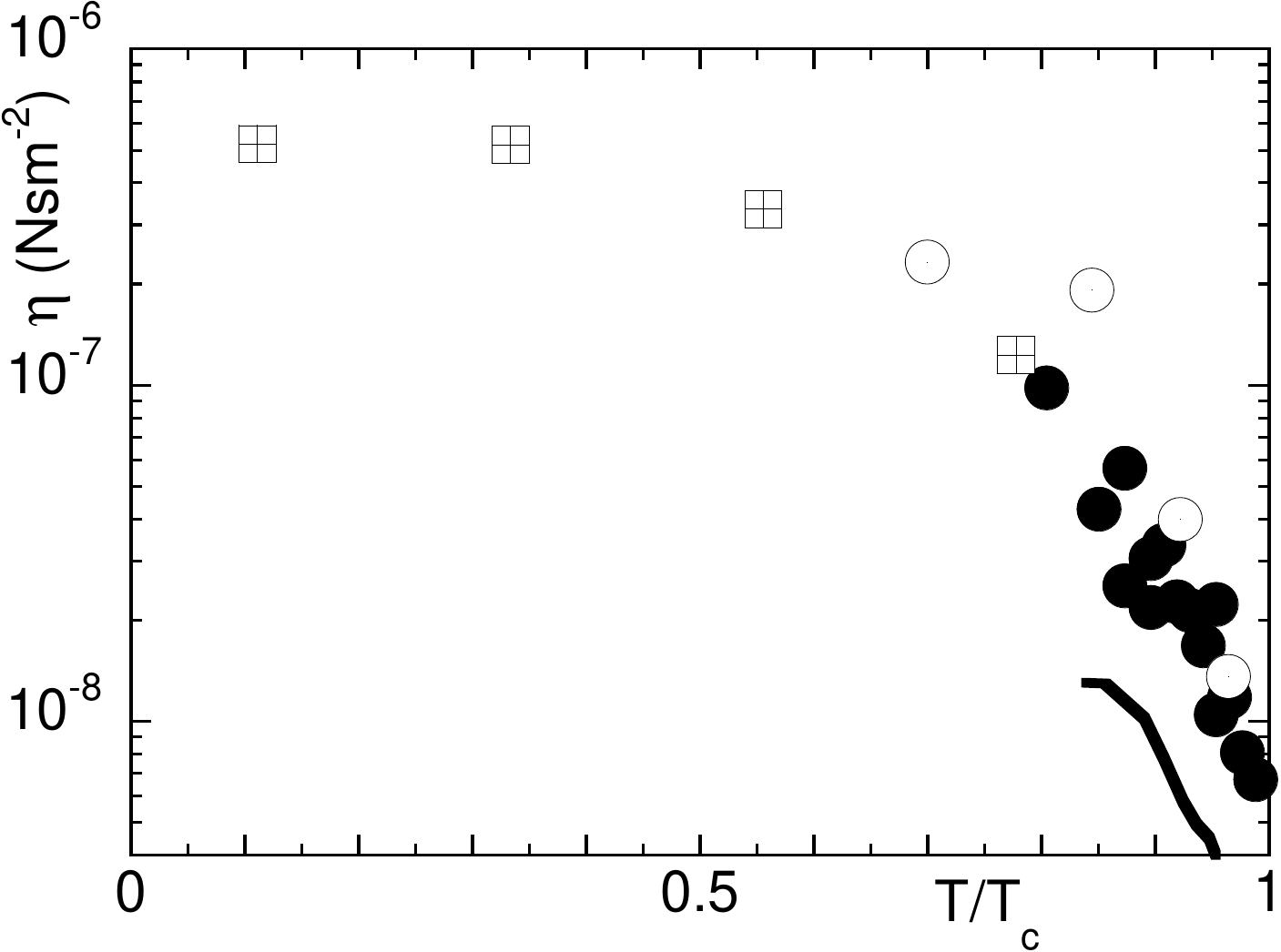}
\\
{Figure 5}
\end{center}
\end{figure}

\end{document}